\documentclass[sigconf]{acmart}

\usepackage[utf8]{inputenc}
\usepackage{tabulary}
\usepackage[english]{babel}
\usepackage[T1]{fontenc}

\usepackage{balance}

\usepackage{subfig}
\usepackage{colortbl}

\usepackage{booktabs}
\usepackage{multirow}
\usepackage{graphicx}
\usepackage{xspace}
\usepackage{adjustbox}

\usepackage[inline]{enumitem}

\usepackage{dsfont}

\theoremstyle{definition}

\usepackage{algorithm}
\usepackage{url}

\usepackage{array}
\newcolumntype{P}[1]{>{\raggedright\arraybackslash}p{#1}}

\usepackage{footmisc} 

\usepackage{tikz}
\usepackage[draft]{tikzpeople}
\usepackage{pgf-umlcd}

\usepackage{forest}
\usetikzlibrary{angles}
\definecolor{drawColor}{RGB}{128 128 128}

\usepackage{listings}

\definecolor{codegreen}{rgb}{0,0.6,0}
\definecolor{codegray}{rgb}{0.5,0.5,0.5}
\definecolor{codepurple}{rgb}{0.58,0,0.82}
\definecolor{backcolour}{rgb}{0.95,0.95,0.92}
\definecolor{litblue}{RGB}{0,132,187}

\lstdefinestyle{UVL}{
  backgroundcolor=\color{backcolour},   
  commentstyle=\color{codegreen},
  keywordstyle=\color{codepurple},
  numberstyle=\tiny\color{codegray},
  stringstyle=\color{litblue},
  basicstyle=\ttfamily\footnotesize,
  breakatwhitespace=false,         
  breaklines=true,                 
  captionpos=b,                    
  keepspaces=true,                 
  numbers=left,                    
  numbersep=5pt,                  
  showspaces=false,                
  showstringspaces=false,
  showtabs=false,                  
  tabsize=2,
  morekeywords={namespace,features,constraints,alternative,or,mandatory,optional, abstract}
}

\newcommand{\travart}{{\fontsize{10}{12}\selectfont T}RA{\fontsize{10}{12}\selectfont V}AR{\fontsize{10}{12}\selectfont T}\xspace}

\acmConference[MODEVAR'24]{6th International Workshop on Languages for Modelling Variability (MODEVAR 2024)}{February 6, 2024}{Bern, Switzerland}


\title{On the Challenges of Transforming UVL to IVML}

\renewcommand\footnotetextcopyrightpermission[1]{}

\author{Prankur Agarwal}
\orcid{0009-0009-5772-6690}
\affiliation{%
  \department{JKU/Dynatrace Co-Innovation Lab, LIT CPS Lab}
  \institution{Dynatrace Research}
  \city{Linz}
   \country{Austria}
}
\email{prankur.agarwal@dynatrace.com}

\author{Kevin Feichtinger}
\orcid{0000-0003-1182-5377}
\affiliation{
  \department{CRC 1608, KASTEL – Dependability of Software-intensive Systems}
  \institution{Karlsruhe Institute of Technology}
  \country{Germany}
}
\email{kevin.feichtinger@kit.edu}

\author{Klaus Schmid}
\orcid{0000-0002-4147-3942 }
\affiliation{
  \department{Software Systems Engineering, Institute of Computer Science}
  \institution{University of Hildesheim}
  \country{Germany}
}
\email{schmid@sse.uni-hildesheim.de}

\author{Holger Eichelberger}
\orcid{0000-0002-2584-5558}
\affiliation{
  \department{Software Systems Engineering, Institute of Computer Science}
  \institution{University of Hildesheim}
  \country{Germany}
}
\email{eichelberger@sse.uni-hildesheim.de}

\author{Rick Rabiser}
\orcid{0000-0003-3862-1112}
\affiliation{
  \department{CDL VaSiCS, LIT CPS Lab}
  \institution{Johannes Kepler University Linz}
  \country{Austria}
}
\email{rick.rabiser@jku.at}

\begin{abstract}
Software product line techniques encourage the reuse and adaptation of software components for creating customized products or software systems. 
These different product variants have commonalities and differences, which are managed by variability modeling. 
Over the past three decades, both academia and industry have developed numerous variability modeling methods, each with its own advantages and disadvantages. 
Many of these methods have demonstrated their utility within specific domains or applications. 
However, comprehending the capabilities and differences among these approaches to pinpoint the most suitable one for a particular use case remains challenging. 
Thus, new modeling techniques and tailored tools for handling variability are frequently created. 
Transitioning between variability models through transformations from different approaches can help in understanding the benefits and drawbacks of different modeling approaches.
However, implementing such transformations presents challenges, such as semantic preservation and avoiding information loss. 
\travart is a tool that helps with transitioning between different approaches by enabling the transformation of variability models into other variability models of different types. 
This paper discusses the challenges for such transformations between UVL and IVML.
It also presents a one-way transformation from the UVL to IVML with as little information loss as possible.

\end{abstract}


\begin{document}
\maketitle

\section{Introduction}
Software Product Line (SPL) engineering encourages the reuse of predefined software artifacts for creating tailored variations of the same software to meet specific customer requirements~\cite{spl_clements, bosch2000design}.
Variability modeling plays a crucial role in SPL engineering~\cite{spl_clements, pohl2005software} as it allows capturing the common and variable characteristics of a set of (software) systems in dedicated models~\cite{berger2013survey}.
These models are then used to derive and customize different software products with varying features and functionalities~\cite{berger2013survey}.

Over the past 30 years, various variability modeling approaches have been developed, each having their own advantages and disadvantages~\cite{berger2013survey,bashroush2017case,chen2011systematic,galster2013variability,Raatikainen2019software,schobbens2006feature}.
The Feature-Oriented Domain Analysis (FODA) approach, introduced by \citet{kang1990feature}, is the basis of most feature modeling approaches available today. 
Another approach is decision modeling, which has been influenced by the Synthesis method~\cite{schmid2011comparison,Synthesis}. 
Even within a single approach, there are multiple variants. 
For example, in feature modeling, one can use academic approaches/tools such as FeatureIDE~\cite{FeatureIDE} or commercial ones such as  pure::variants~\cite{puresystemsUserguide2022}. 
Similarly, in decision modeling, one can use DOPLER~\cite{DOPLER} or Integrated Variability Modeling Language (IVML) \cite{EichelbergerSchmid14c,ivml_dsl}.
Beyond these common approaches~\cite{CzarneckiCool2012} also textual variability modeling languages~\cite{terBeek2019textualoverview}, Orthogonal Variability Modeling (OVM)~\cite{pohl2005software}, Unified Modeling Language (UML)-based variability modeling~\cite{gomaa2005designing}, and the Common Variability Language (CVL)~\cite{haugen2013cvl} are available. 
Additionally, various open-source communities and industries have developed their own solutions to model variability. 
For instance, the KConfig language supporting configuration of the Linux Kernel~\cite{she2010variability} and the Component Definition Language (CDL)~\cite{berger2010variability} from the eCos~\cite{veer2011ecos} operating system. 
Despite numerous efforts, such as the CVL~\cite{haugen2013cvl} or Universal Variability Language (UVL)~\cite{SundermannUVLSPLC2021}, there is no officially accepted standard in the variability modeling community, today. 

The lack of standardization has led to the need for tools that support multiple variability modeling languages through importers and exporters. 
However, most of these approaches and tools have a limited lifespan~\cite{bashroush2017case,berger2019usage,beyond_spl,spl_achievement_challenge}. 
As a result, researchers and practitioners often resort to creating new approaches instead of exploring existing ones~\cite{kevin_thesis} and potentially re-using, customizing, and integrating them. 
However, various  differences among approaches make it hard for researchers and practitioners to compare their advantages and disadvantages.
Thus, it would be advantageous to concentrate on enhancing the interoperability of current variability modeling tools~\cite{spl_achievement_challenge} to choose the most suitable approach for a given use case.
\travart~\cite{FeichtingerTraVarTVaMoS2021} was developed to increase the interoperability of existing variability modeling tools by transforming different variability models into each other while preserving the variability to its maximum by minimizing information loss~\cite{feichtingerSPLC2022informationloss}. 
\travart~\cite{FeichtingerTraVarTVaMoS2021} uses UVL~\cite{SundermannUVLSPLC2021} as a pivot language and already supports various variability modeling languages like FeatureIDE feature models~\cite{FeatureIDE}, DOPLER~\cite{DOPLER}, OVM~\cite{pohl2005software}, Product-Process-Resource Domain-Specific Language (PPR-DSL)~\cite{Meixner2021a,MeixnerVaMos2022EfficientProcessExploration}, and\\pure::variants~\cite{puresystemsUserguide2022,RomanoMODEVAR2022}. 
However, there are many more languages that are currently not supported, such as the IVML~\cite{EichelbergerSchmid14c}.

In this paper, we discuss the differences between UVL~\cite{SundermannUVLSPLC2021} and IVML~\cite{EichelbergerSchmid14c} and the resulting challenges for  their transformation.
We also describe the process of deriving and developing transformations from UVL~\cite{SundermannUVLSPLC2021} to IVML~\cite{EichelbergerSchmid14c} using \travart~\cite{FeichtingerTraVarTVaMoS2021}.

\section{Background}
In this section, we discuss relevant background and related work to build transformations between UVL~\cite{SundermannUVLSPLC2021} and IVML~\cite{EichelbergerSchmid14c} using an extended Onlineshop case study~\cite{online_shop_case_study, kevin_thesis}.

\subsection{Universal Variability Language}
The UVL is a community effort towards a unified language for variability models developed and maintained by the MODEVAR initiative\footnote{MODEVAR initiative – \href{https://modevar.github.io/}{https://modevar.github.io/}}~\cite{BenavidesMODEVAR2019}. UVL is a feature modeling approach that should allow easy access to datasets and analyses used by other researchers.
Listing~\ref{lst:uvl_online_shop} shows the UVL model for the extended Onlineshop~\cite{online_shop_case_study, kevin_thesis} case study.
A feature can be \textit{abstract} (no implementation and for grouping purposes, e.g., \texttt{Payment} in line 6) or \textit{concrete} (actual implementations of the feature, e.g., \texttt{Catalog} in line 12). 
Furthermore, a feature can be \textit{mandatory} (e.g., \texttt{UserManagement} in line 18) or \textit{optional} (e.g., \texttt{Search} in line 17). 
Among features of the feature model, relations exist, like features of a feature group can be \textit{alternatives} (meaning only one of the feature can be selected, e.g., \texttt{DebitCard} and \texttt{CreditCard} in lines 8 and 9) or an \textit{or} group (meaning at least one of the features must be selected, e.g., \texttt{Mobile}, \texttt{Tablet} and \texttt{PC} in lines 38-40). 
Additionally, features can depend on each other, which is modeled using constraints (cf. lines 42-45).

Different usage scenarios often require different variability languages. 
However, adding more language features to UVL can complicate its integration into existing tools. 
Thus, UVL utilizes language levels~\cite{ThuemLanguageLevelsFM2019} for more advanced language features~\cite{uvl_levels}. 
UVL currently supports three language levels - Boolean, Arithmetic, and Type. 
The Type level supports basic datatypes like String, Boolean, Integer, and Real.

\begin{lstlisting}[caption=UVL~\cite{SundermannUVLSPLC2021} Model for the extended Onlineshop case study~\cite{online_shop_case_study, kevin_thesis}.,label={lst:uvl_online_shop},style=UVL]
namespace Onlineshop

features
	Onlineshop {abstract true}
            mandatory
                Payment {abstract true}
                    alternative
                        DebitCard
                        CreditCard
                ProductSelection {abstract true}
                    mandatory
                        Catalog
                            optional
                                    Categories
                                    Sort
                    optional
                        Search
                UserManagement {abstract true}
                    or
                        Orders
                        Security
                        Payments
                        Wishlist
            optional
                Newsletter
            mandatory
                Design
                    mandatory
                        Responsive
                        Review 
                            [2..3]
                                Stars
                                Numerical
                                Comments
                ShoppingBasket
                Platform
                    or
                        Mobile
                        Tablet
                        PC
constraints
    Sort | Search
    Search => Security
    Payments => !Security
    Security => !Payments
\end{lstlisting}

\subsection{Integrated Variability Modeling Language}
IVML~\cite{EichelbergerSchmid14c} is a text-based variability modeling language that helps describe configurations of variability-rich software ecosystems~\cite{ivml_dsl}. 
IVML is implemented as part of
EASY-Producer~\cite{Eichelberger2014EASyProducer,ivml_easy_producer}, a toolset for creating and transforming product lines defined by IVML models more effectively.
IVML unifies variability modeling and configuration into a single language, remains implementation-agnostic, and employs strong typing with common types like Integer and Boolean.
IVML is divided into two parts: a core modeling language and an advanced modeling language that extends the core\\language~\cite{ivml_dsl, ivmlSpecification2015}. 

One of the core concepts in IVML is to represent variability as a typed variable~\cite{ivml_dsl}.
Boolean variables are utilized to represent optional features, user-defined enumerations capture alternatives, and collections model multiple selections~\cite{ivmlSpecification2015}. 
Constraints, similar to those seen in Object Constraint Language (OCL), including relational expressions and quantifiers, are imposed to effectively shape and confine the permissible selections. 
IVML also introduces the concept of compounds to encapsulate various forms of variabilities. 
A compound can be used to characterize an individual variability, a multiple selection within a collection, or even for nesting feature modeling to construct a hierarchical structure. 
Within IVML, variabilities can be refined, restricted, or shared through their types. 
Compound types can inherit variables from their parent types and add new ones through type refinement, which is similar to inheritance in object-oriented languages. 
On the other hand, type restriction defines new types by constraining existing ones, allowing for more specialized variability types. 
Sharing is another powerful capability that allows variables to reference each other, with typed references ensuring that valid target types are maintained, even in the presence of type refinements. 
Value bindings are created by using either assignment or value propagation constraints, and variables can have default values to simplify configurations. 
IVML annotations also support default values and constraints, making the modeling language more expressive.

Listing~\ref{lst:ivml_online_shop} shows an IVML model for the extended Onlineshop case study~\cite{online_shop_case_study, kevin_thesis}.
It contains enumerations (cf. Lines 2, 8, 12, and 15) and sets (cf. Lines 9, 13, and 16) to showcase the variability. 
The \texttt{size} function is used to define the cardinality (cf. Lines 10, 14, and 17) whereas \texttt{isDefined} (cf. Line 4) is used to define a mandatory feature.
Line 18 adds propositional logic constraints and lines 19-21 define implication constraints.

\begin{lstlisting}[caption=IVML~\cite{EichelbergerSchmid14c} Model for the extended Onlineshop case study~\cite{online_shop_case_study,kevin_thesis}.,label={lst:ivml_online_shop},morekeywords={Boolean, Integer,enum, isDefined,setOf, implies, size, project,includes}]
project OnlineShop {
    enum PaymentTypes {DebitCard, CreditCard};
    PaymentTypes Payment;
    isDefined(Payment);
    Boolean Categories;
    Boolean Sort;
    Boolean Search;
    enum UserManagementOptions {Orders, Security, Payments, Wishlist};
    setOf(UserManagementOptions) UserManagement;
    size(UserManagement) >= 1;
    Boolean Newsletter;
    enum ReviewTypes {Stars, Numerical, Comments};
    setOf(ReviewTypes) Review;
    size(Review) >= 2;
    enum PlatformType {Mobile, Tablet, PC};
    setOf(PlatformType) Platform;
    size(Platform) >= 1;
    Sort or Search;
    Search implies includes(UserManagement, UserManagementOptions.Security);
    includes(UserManagement, UserManagementOptions.Payments) implies (includes(UserManagement, UserManagementOptions.Security) <> true);
    includes(UserManagement, UserManagementOptions.Security) implies (includes(UserManagement, UserManagementOptions.Payments) <> true);
}
\end{lstlisting}

\section{IVML vs. UVL: Comparison and Transformation Challenges}
\label{sec:comparision}
UVL~\cite{SundermannUVLSPLC2021} and IVML~\cite{EichelbergerSchmid14c} are two different languages for describing variability models in SPL. 
They have some similarities, but they also differ in how they express syntax, semantics, and features. 
This section compares these two languages in detail, focusing on explaining the essential transformation challenges. 

UVL is a result of a collaborative effort to create a standard format for variability modeling~\cite{SundermannUVLSPLC2021} by the MODEVAR initiative~\cite{BenavidesMODEVAR2019}. 
IVML is a language initially created in the FP7 INDENICA project to handle variability in service-based systems~\cite{EichelbergerSchmid14c, ivml_dsl}. 
UVL is based on the idea of a feature model, which is a tree-like representation of common and variable features. 
A feature model has a root feature, which represents the product line, and a set of sub-features, which represent options for the product line. 
UVL uses indentation to define feature hierarchy and keywords such as mandatory, optional, alternative, and or to define feature cardinality and variability. 
On the contrary, IVML is based on the idea of characterizing the variability space in a more general manner and defining specific configurations in the same language. 
An IVML model consists of declarations that define variability decisions, variability kinds through types, attributes, and constraints. 
IVML uses a syntax similar to Java or C\# and constraints to declaratively define relationships between features and their parent features or to assign decision values. 
In their simplest forms UVL does not support data types, while IVML does.

UVL only supports basic constraints with propositional logic~\cite{SundermannUVLSPLC2021, uvl_levels}, while IVML offers more expressive power with arithmetic expressions, assignments, and attribute definitions up to first-order logic~\cite{EichelbergerSchmid14c, ivml_dsl}. 
UVL is integrated with various software tools, such as FeatureIDE~\cite{FeatureIDE} and \travart~\cite{FeichtingerTraVarTVaMoS2021}. 
IVML, however, is mainly used by the EASy-Producer tool suite~\cite{Eichelberger2014EASyProducer,ivml_easy_producer}.

In summary, UVL and IVML have different objectives.
UVL is designed as a universal language that can express the core of any feature model. 
Its aim is to offer a standard format that enables the sharing and compatibility of feature models among various tools and platforms. 
IVML is intentionally created as a Domain Specific Language (DSL), customized to address the specific challenges and demands of service-oriented systems. 
It acts as a powerful language that can handle complex variability modeling tasks such as analysis, configuration, adaptation, and evolution.

Transforming between UVL and IVML is challenging due to their differences, especially in their constraint languages.
IVML is more expressive than UVL. 
UVL can only handle a limited set of constraints that IVML supports. 
For instance, UVL does not allow constraints such as \texttt{if}, \texttt{def}, etc. 
Furthermore, UVL does not have the advanced modeling features of IVML such as \texttt{assign}, \texttt{conflicts}, etc. 
This implies that a conversion from UVL to IVML is possible (but some concepts need to be adjusted, especially regarding structure), while a conversion from IVML to UVL will necessarily result in losing information for more complex IVML models.

\section{Initial Transformations}
In this section, we provide a detailed description of the methods used to perform a one-way transformation from UVL~\cite{SundermannUVLSPLC2021} to\\IVML~\cite{EichelbergerSchmid14c}, taking the differences between the two approaches into account (cf. Section~\ref{sec:comparision}).
Building on earlier work by~\citet{ElSharkawyDederichsSchmid12a}, the main objective of this transformation is to preserve the variability as possible while ignoring structural losses~\cite{feichtingerSPLC2022informationloss,kevin_thesis}. 
This is because, even when the new modeling elements are added, the fundamental goal still is to protect the configuration space.

During the transformation, we use special names for newly introduced model elements.
These special names help maintain name uniqueness and allow distinguishing between old and new model elements for readability.
By using these special names, we try to create a system that everyone can understand and work with easily.
The suffixes used to create these special names are:
\begin{itemize}
    \item Enum Declaration: \texttt{\_\_ENUM\_\_<number>}
    \item Enum Instance Declaration:\\ \texttt{\_\_ENUM\_\_<number>\_\_INSTANCE}
    \item Set Declaration: \texttt{\_\_SET\_\_<number>}
    \item Set Instance Declaration: \texttt{\_\_SET\_\_<number>\_\_INSTANCE}
    \item Compound Declaration: \texttt{\_\_COMPOUND\_\_<number>}
    \item Compound Instance Declaration: \\ \texttt{\_\_COMPOUND\_\_<number>\_\_INSTANCE}
\end{itemize}

In these conventions, the \textit{number} is the number (starting from~1) of the \texttt{Enum}, \texttt{Set}, or \texttt{Compound} created from the same parent.

The one-way transformation from UVL to IVML mainly focuses on three concepts of UVL - features, or group, and alternative group.
If it is a part of an or-group, or an alternative-group, the feature is either mandatory or optional. 
Then it is directly transformed into a \texttt{Boolean} variable. 
For mandatory features, in the target model, an additional \texttt{isDefined} constraint is added to the created model during the transformation (cf. Lines 5-7 in Listing~\ref{lst:ivml_online_shop}, which is without suffixes due to space constraints). 
Moreover, the mandatory features directly from the root feature or with all the parents as mandatory features, are ignored during transformation as they do not contribute to the variability. 
For instance, feature \texttt{Catalog} in Listing~\ref{lst:uvl_online_shop} is not transformed and hence is not present in Listing~\ref{lst:ivml_online_shop}.
An alternative group is transformed into an enumeration and then this enumeration is instantiated. 
Further constraints can be added to this created instance of the enumeration depending on the parent feature type of the group. 
The constraint is of the form \\\texttt{<parent\_inclusion\_condition> implies isDefined(\\enumeration\_instance)} (cf. Lines 2-4 in Listing~\ref{lst:ivml_online_shop}).
Similarly, the or group is transformed into an enumeration but instead of creating an instance of this enumeration, a set is defined of this enumeration.
This set has all the possible combinations of choices from this group, including none and all. 
Lastly, we add a constraint for the size of this set to be at least 1.
The constraint is of the form \texttt{<parent\_inclusion\_condition> implies size(set) >= 1} (cf. Lines 8-10 in Listing~\ref{lst:ivml_online_shop}).


The next step after transforming all the features is to transform the constraints. 
Table~\ref{tab:constraint_mapping} shows the mapping of constraints. 
The table reveals that the transformation of constraints is simple and straightforward, and it is like a one-to-one mapping between the two languages. 
For example, the constraint \texttt{Sort | Search} in line 42 of Listing~\ref{lst:uvl_online_shop} is directly transformed to \texttt{Sort or Search} as shown in line 18 of Listing~\ref{lst:ivml_online_shop}.
Further, the implication constraint, \texttt{Search => Security} in line 43 of Listing~\ref{lst:uvl_online_shop} is transformed to \texttt{Search implies includes(UserManagement,\\UserManagementOptions.Security)} as shown in line 19 of Listing~\ref{lst:ivml_online_shop}.
It is important to note the type of the variable during the transformation of constraints. 
For instance, the \texttt{includes} keyword is used for checking whether a particular value exists in a set (cf. Lines 20-21 in Listing~\ref{lst:ivml_online_shop}).
This simplicity in constraint transformation helps in maintaining the variability and hence prevents information loss.

\begin{table}[ht!]
\centering
\scriptsize
\caption{Constraint mappings for UVL~\cite{SundermannUVLSPLC2021} and IVML~\cite{EichelbergerSchmid14c}.}
\label{tab:constraint_mapping}
\begin{adjustbox}{max width=\textwidth}
\scriptsize
{\renewcommand{\arraystretch}{2}%
\begin{tabular}{|c|c|}
\hline
\multicolumn{1}{|c|}{\textbf{UVL}} & \multicolumn{1}{c|}{\textbf{IVML}} \\
    \hline
         and / or / not &and / or / not\\
         \rowcolor{codegray!8}iff&iff\\
        implies&implies\\
        \rowcolor{codegray!8}len(<string\_feature>) & size(<string\_variable>)\\
         floor(<numeric\_feature>)&floor(<numeric\_variable>)\\
         \rowcolor{codegray!8} > / >= / < / <= &> / >= / < / <=\\
         == / != &== / !=\\
         \rowcolor{codegray!8}add expression&add\\
         subtract expression&subtract\\
         \rowcolor{codegray!8}multiplication expression&multiplication\\
         division expression&division\\
    \hline
\end{tabular}%
}
\end{adjustbox}
\end{table}

Currently, the verification of the transformed model is restricted to its syntax correctness, using the IVML parser of the EASy-\\Producer~\cite{Eichelberger2014EASyProducer, ivml_easy_producer} tool. 
We aim to extend that in future work.

\section{Conclusion and Research Agenda}

In this paper, we described the challenges of transforming UVL and IVML models.
We also explored a way to perform a one-way transformation of UVL models into IVML models.
We mapped UVL concepts to their IVML equivalents with a focus on avoiding any loss of information in terms of variability and semantics. 
Our mapping strategies also prioritized human readability while transforming the languages. 

In future work, we aim to further investigate the mappings of the concepts in these languages and also implement the one-way transformation from IVML to UVL, and round-trip transformations from both IVML and UVL.
Moreover, we plan to work on verification methods for the transformed models as well as on optimizing the implementation of the transformation.
These verification methods will include but are not limited to, generating the valid and invalid configurations of the original UVL model and applying them to the transformed IVML model.
We will integrate these transformations in the existing transformation tool \travart.
Lastly, the support for model evolution has yet to be investigated for IVML based on preliminary work~\cite{FeichtingerMBR22}.

\begin{acks}
The financial support by the Christian Doppler Research Association, the Austrian Federal Ministry for Digital and Economic Affairs and the National Foundation for Research, Technology and Development is gratefully acknowledged. Partially funded by the Deutsche Forschungsgemeinschaft (DFG, German Research Foundation) – CRC 1608 – 501798263. This work is partially supported by the German Federal Ministry of Economic Affairs and Climate Action (BMWK, IIP-Ecosphere -- 01MK20006C).
\end{acks}

\balance
\bibliographystyle{ACM-Reference-Format}
\bibliography{references}


\begin{thebibliography}{43}


\ifx \showCODEN    \undefined \def \showCODEN     #1{\unskip}     \fi
\ifx \showDOI      \undefined \def \showDOI       #1{#1}\fi
\ifx \showISBNx    \undefined \def \showISBNx     #1{\unskip}     \fi
\ifx \showISBNxiii \undefined \def \showISBNxiii  #1{\unskip}     \fi
\ifx \showISSN     \undefined \def \showISSN      #1{\unskip}     \fi
\ifx \showLCCN     \undefined \def \showLCCN      #1{\unskip}     \fi
\ifx \shownote     \undefined \def \shownote      #1{#1}          \fi
\ifx \showarticletitle \undefined \def \showarticletitle #1{#1}   \fi
\ifx \showURL      \undefined \def \showURL       {\relax}        \fi
\providecommand\bibfield[2]{#2}
\providecommand\bibinfo[2]{#2}
\providecommand\natexlab[1]{#1}
\providecommand\showeprint[2][]{arXiv:#2}

\bibitem[\protect\citeauthoryear{Arif, de~Boer, Helvensteijn, Villela, and
  Wong}{Arif et~al\mbox{.}}{2012}]%
        {online_shop_case_study}
\bibfield{author}{\bibinfo{person}{Taslim Arif}, \bibinfo{person}{Frank de
  Boer}, \bibinfo{person}{Michiel Helvensteijn}, \bibinfo{person}{Karina
  Villela}, {and} \bibinfo{person}{Peter Wong}.}
  \bibinfo{year}{2012}\natexlab{}.
\newblock \bibinfo{title}{Evaluation of Modeling. Deliverable 5.3 of project
  FP7-231620 (HATS). Website}.
\newblock
\newblock
\urldef\tempurl%
\url{http://www.hats-project.eu/sites/default/files/Deliverable5.3.pdf#page=55.
  2012}
\showURL{%
\tempurl}


\bibitem[\protect\citeauthoryear{Bashroush, Garba, Rabiser, Groher, and
  Botterweck}{Bashroush et~al\mbox{.}}{2017}]%
        {bashroush2017case}
\bibfield{author}{\bibinfo{person}{Rabih Bashroush}, \bibinfo{person}{Muhammad
  Garba}, \bibinfo{person}{Rick Rabiser}, \bibinfo{person}{Iris Groher}, {and}
  \bibinfo{person}{Goetz Botterweck}.} \bibinfo{year}{2017}\natexlab{}.
\newblock \showarticletitle{Case tool support for variability management in
  software product lines}.
\newblock \bibinfo{journal}{\emph{ACM Computing Surveys (CSUR)}}
  \bibinfo{volume}{50}, \bibinfo{number}{1} (\bibinfo{year}{2017}),
  \bibinfo{pages}{14:1--14:45}.
\newblock


\bibitem[\protect\citeauthoryear{Beek, Schmid, and Eichelberger}{Beek
  et~al\mbox{.}}{2019}]%
        {terBeek2019textualoverview}
\bibfield{author}{\bibinfo{person}{Maurice H.~ter Beek}, \bibinfo{person}{Klaus
  Schmid}, {and} \bibinfo{person}{Holger Eichelberger}.}
  \bibinfo{year}{2019}\natexlab{}.
\newblock \showarticletitle{{Textual Variability Modeling Languages: An
  Overview and Considerations}}. In \bibinfo{booktitle}{\emph{Proceedings of
  the 23rd International Systems and Software Product Line Conference - Volume
  B}} \emph{(\bibinfo{series}{SPLC '19})}. \bibinfo{publisher}{ACM},
  \bibinfo{address}{New York, NY, USA}, \bibinfo{pages}{151–157}.
\newblock
\showISBNx{9781450366687}


\bibitem[\protect\citeauthoryear{Benavides, Rabiser, Batory, and
  Acher}{Benavides et~al\mbox{.}}{2019}]%
        {BenavidesMODEVAR2019}
\bibfield{author}{\bibinfo{person}{David Benavides}, \bibinfo{person}{Rick
  Rabiser}, \bibinfo{person}{Don Batory}, {and} \bibinfo{person}{Mathieu
  Acher}.} \bibinfo{year}{2019}\natexlab{}.
\newblock \showarticletitle{First International Workshop on Languages for
  Modelling Variability (MODEVAR 2019)}. In
  \bibinfo{booktitle}{\emph{Proceedings of the 23rd International Systems and
  Software Product Line Conference - Volume A}} \emph{(\bibinfo{series}{SPLC
  '19})}. \bibinfo{publisher}{ACM}, \bibinfo{address}{New York, NY, USA},
  \bibinfo{pages}{323}.
\newblock
\showISBNx{9781450371384}


\bibitem[\protect\citeauthoryear{Berger and Collet}{Berger and Collet}{2019}]%
        {berger2019usage}
\bibfield{author}{\bibinfo{person}{Thorsten Berger} {and}
  \bibinfo{person}{Philippe Collet}.} \bibinfo{year}{2019}\natexlab{}.
\newblock \showarticletitle{Usage Scenarios for a Common Feature Modeling
  Language}. In \bibinfo{booktitle}{\emph{Proceedings of the 23rd International
  Systems and Software Product Line Conference - Volume B}}
  \emph{(\bibinfo{series}{SPLC '19})}. \bibinfo{publisher}{ACM},
  \bibinfo{address}{New York, NY, USA}, \bibinfo{pages}{174–181}.
\newblock
\showISBNx{9781450366687}


\bibitem[\protect\citeauthoryear{Berger, Rublack, Nair, Atlee, Becker,
  Czarnecki, and W{\k{a}}sowski}{Berger et~al\mbox{.}}{2013}]%
        {berger2013survey}
\bibfield{author}{\bibinfo{person}{Thorsten Berger}, \bibinfo{person}{Ralf
  Rublack}, \bibinfo{person}{Divya Nair}, \bibinfo{person}{Joanne~M Atlee},
  \bibinfo{person}{Martin Becker}, \bibinfo{person}{Krzysztof Czarnecki}, {and}
  \bibinfo{person}{Andrzej W{\k{a}}sowski}.} \bibinfo{year}{2013}\natexlab{}.
\newblock \showarticletitle{A survey of variability modeling in industrial
  practice}. In \bibinfo{booktitle}{\emph{Proc. of the 7th International
  Workshop on Variability Modelling of Software-intensive Systems}}. ACM,
  \bibinfo{pages}{7--14}.
\newblock


\bibitem[\protect\citeauthoryear{Berger, She, Lotufo, W{\k{a}}sowski, and
  Czarnecki}{Berger et~al\mbox{.}}{2010}]%
        {berger2010variability}
\bibfield{author}{\bibinfo{person}{Thorsten Berger}, \bibinfo{person}{Steven
  She}, \bibinfo{person}{Rafael Lotufo}, \bibinfo{person}{Andrzej
  W{\k{a}}sowski}, {and} \bibinfo{person}{Krzysztof Czarnecki}.}
  \bibinfo{year}{2010}\natexlab{}.
\newblock \showarticletitle{Variability modeling in the real: a perspective
  from the operating systems domain}. In \bibinfo{booktitle}{\emph{Proc. of the
  IEEE/ACM International Conference on Automated Software Engineering}}. ACM,
  \bibinfo{pages}{73--82}.
\newblock


\bibitem[\protect\citeauthoryear{Bosch}{Bosch}{2000}]%
        {bosch2000design}
\bibfield{author}{\bibinfo{person}{J. Bosch}.} \bibinfo{year}{2000}\natexlab{}.
\newblock \bibinfo{booktitle}{\emph{Design and Use of Software Architectures:
  Adopting and Evolving a Product-line Approach}}.
\newblock \bibinfo{publisher}{Addison-Wesley}.
\newblock
\showISBNx{9780201674941}
\showLCCN{00036362}
\urldef\tempurl%
\url{https://books.google.at/books?id=FDfyWknLvMYC}
\showURL{%
\tempurl}


\bibitem[\protect\citeauthoryear{Chen and Babar}{Chen and Babar}{2011}]%
        {chen2011systematic}
\bibfield{author}{\bibinfo{person}{Lianping Chen} {and}
  \bibinfo{person}{Muhammad~Ali Babar}.} \bibinfo{year}{2011}\natexlab{}.
\newblock \showarticletitle{A systematic review of evaluation of variability
  management approaches in software product lines}.
\newblock \bibinfo{journal}{\emph{Information and Software Technology}}
  \bibinfo{volume}{53}, \bibinfo{number}{4} (\bibinfo{year}{2011}),
  \bibinfo{pages}{344--362}.
\newblock


\bibitem[\protect\citeauthoryear{Clements and Northrop}{Clements and
  Northrop}{2002}]%
        {spl_clements}
\bibfield{author}{\bibinfo{person}{P. Clements} {and} \bibinfo{person}{L.
  Northrop}.} \bibinfo{year}{2002}\natexlab{}.
\newblock \bibinfo{booktitle}{\emph{Software Product Lines: Practices and
  Patterns}}.
\newblock \bibinfo{publisher}{Addison-Wesley}.
\newblock
\showISBNx{9780201703320}
\showLCCN{2001035285}


\bibitem[\protect\citeauthoryear{Consortium}{Consortium}{1991}]%
        {Synthesis}
\bibfield{author}{\bibinfo{person}{Software~Productivity Consortium}.}
  \bibinfo{year}{1991}\natexlab{}.
\newblock \bibinfo{booktitle}{\emph{{Synthesis Guidebook}}}.
\newblock \bibinfo{type}{{T}echnical {R}eport}.
  \bibinfo{institution}{SPC-91122-MC. Herndon, Virginia: Software Productivity
  Consortium}.
\newblock


\bibitem[\protect\citeauthoryear{Czarnecki, Grünbacher, Rabiser, Schmid, and
  W{\k{a}}sowski}{Czarnecki et~al\mbox{.}}{2012}]%
        {CzarneckiCool2012}
\bibfield{author}{\bibinfo{person}{Krzysztof Czarnecki}, \bibinfo{person}{Paul
  Grünbacher}, \bibinfo{person}{Rick Rabiser}, \bibinfo{person}{Klaus Schmid},
  {and} \bibinfo{person}{Andrzej W{\k{a}}sowski}.}
  \bibinfo{year}{2012}\natexlab{}.
\newblock \showarticletitle{{Cool Features and Tough Decisions: A Comparison of
  Variability Modeling Approaches}}. In \bibinfo{booktitle}{\emph{Proc. of the
  6th International Workshop on Variability Modeling of Software-Intensive
  Systems}}. \bibinfo{publisher}{ACM}, \bibinfo{pages}{173--182}.
\newblock


\bibitem[\protect\citeauthoryear{Dhungana, Grünbacher, and Rabiser}{Dhungana
  et~al\mbox{.}}{2011}]%
        {DOPLER}
\bibfield{author}{\bibinfo{person}{Deepak Dhungana}, \bibinfo{person}{Paul
  Grünbacher}, {and} \bibinfo{person}{Rick Rabiser}.}
  \bibinfo{year}{2011}\natexlab{}.
\newblock \showarticletitle{{The DOPLER Meta-Tool for Decision-Oriented
  Variability Modeling: A Multiple Case Study}}.
\newblock \bibinfo{journal}{\emph{Automated Software Engineering}}
  \bibinfo{volume}{18}, \bibinfo{number}{1} (\bibinfo{year}{2011}),
  \bibinfo{pages}{77--114}.
\newblock


\bibitem[\protect\citeauthoryear{Eichelberger, El-Sharkawy, Kr\"{o}her, and
  Schmid}{Eichelberger et~al\mbox{.}}{2014}]%
        {Eichelberger2014EASyProducer}
\bibfield{author}{\bibinfo{person}{Holger Eichelberger},
  \bibinfo{person}{Sascha El-Sharkawy}, \bibinfo{person}{Christian Kr\"{o}her},
  {and} \bibinfo{person}{Klaus Schmid}.} \bibinfo{year}{2014}\natexlab{}.
\newblock \showarticletitle{EASy-Producer: Product Line Development for
  Variant-Rich Ecosystems}. In \bibinfo{booktitle}{\emph{Proceedings of the
  18th International Software Product Line Conference: Companion Volume for
  Workshops, Demonstrations and Tools - Volume 2}} \emph{(\bibinfo{series}{SPLC
  '14})}. \bibinfo{publisher}{Association for Computing Machinery},
  \bibinfo{address}{New York, NY, USA}, \bibinfo{pages}{133–137}.
\newblock
\showISBNx{9781450327398}


\bibitem[\protect\citeauthoryear{Eichelberger and Schmid}{Eichelberger and
  Schmid}{2015a}]%
        {ivml_dsl}
\bibfield{author}{\bibinfo{person}{Holger Eichelberger} {and}
  \bibinfo{person}{Klaus Schmid}.} \bibinfo{year}{2015}\natexlab{a}.
\newblock \showarticletitle{IVML: A DSL for Configuration in Variability-Rich
  Software Ecosystems}. In \bibinfo{booktitle}{\emph{Proceedings of the 19th
  International Conference on Software Product Line}}
  \emph{(\bibinfo{series}{SPLC '15})}. \bibinfo{publisher}{Association for
  Computing Machinery}, \bibinfo{address}{New York, NY, USA},
  \bibinfo{pages}{365–369}.
\newblock


\bibitem[\protect\citeauthoryear{Eichelberger and Schmid}{Eichelberger and
  Schmid}{2015b}]%
        {EichelbergerSchmid14c}
\bibfield{author}{\bibinfo{person}{Holger Eichelberger} {and}
  \bibinfo{person}{Klaus Schmid}.} \bibinfo{year}{2015}\natexlab{b}.
\newblock \showarticletitle{Mapping the Design-Space of Textual Variability
  Modeling Languages: A Refined Analysis}.
\newblock \bibinfo{journal}{\emph{International Journal of Software Tools for
  Technology Transfer}} \bibinfo{volume}{17}, \bibinfo{number}{5}
  (\bibinfo{year}{2015}), \bibinfo{pages}{559--584}.
\newblock


\bibitem[\protect\citeauthoryear{El-Sharkawy, Dederichs, and
  Schmid}{El-Sharkawy et~al\mbox{.}}{2012}]%
        {ElSharkawyDederichsSchmid12a}
\bibfield{author}{\bibinfo{person}{Sascha El-Sharkawy},
  \bibinfo{person}{Stephan Dederichs}, {and} \bibinfo{person}{Klaus Schmid}.}
  \bibinfo{year}{2012}\natexlab{}.
\newblock \showarticletitle{From Feature Models to Decision Models and Back
  Again: An Analysis Based on Formal Transformations}. In
  \bibinfo{booktitle}{\emph{Proc. of the 16th International Software Product
  Line Conference}}. \bibinfo{publisher}{ACM}, \bibinfo{pages}{126--135}.
\newblock


\bibitem[\protect\citeauthoryear{Feichtinger}{Feichtinger}{2023}]%
        {kevin_thesis}
\bibfield{author}{\bibinfo{person}{Kevin Feichtinger}.}
  \bibinfo{year}{2023}\natexlab{}.
\newblock \bibinfo{title}{A Flexible Approach For Transforming Variability
  Artifacts}.
\newblock
\newblock
\urldef\tempurl%
\url{https://resolver.obvsg.at/urn:nbn:at:at-ubl:1-66213}
\showURL{%
\tempurl}


\bibitem[\protect\citeauthoryear{Feichtinger, Meixner, Biffl, and
  Rabiser}{Feichtinger et~al\mbox{.}}{2022a}]%
        {FeichtingerMBR22}
\bibfield{author}{\bibinfo{person}{Kevin Feichtinger}, \bibinfo{person}{Kristof
  Meixner}, \bibinfo{person}{Stefan Biffl}, {and} \bibinfo{person}{Rick
  Rabiser}.} \bibinfo{year}{2022}\natexlab{a}.
\newblock \showarticletitle{Evolution Support for Custom Variability Artifacts
  Using Feature Models: {A} Study in the Cyber-Physical Production Systems
  Domain}. In \bibinfo{booktitle}{\emph{Reuse and Software Quality - 20th
  International Conference on Software and Systems Reuse, {ICSR} 2022,
  Montpellier, France, June 15-17, 2022, Proceedings}}
  \emph{(\bibinfo{series}{Lecture Notes in Computer Science})},
  \bibfield{editor}{\bibinfo{person}{Gilles Perrouin}, \bibinfo{person}{Naouel
  Moha}, {and} \bibinfo{person}{Abdelhak{-}Djamel Seriai}} (Eds.),
  Vol.~\bibinfo{volume}{13297}. \bibinfo{publisher}{Springer},
  \bibinfo{pages}{79--84}.
\newblock
\urldef\tempurl%
\url{https://doi.org/10.1007/978-3-031-08129-3\_5}
\showDOI{\tempurl}


\bibitem[\protect\citeauthoryear{Feichtinger, St\"{o}bich, Romano, and
  Rabiser}{Feichtinger et~al\mbox{.}}{2021}]%
        {FeichtingerTraVarTVaMoS2021}
\bibfield{author}{\bibinfo{person}{Kevin Feichtinger}, \bibinfo{person}{Johann
  St\"{o}bich}, \bibinfo{person}{Dario Romano}, {and} \bibinfo{person}{Rick
  Rabiser}.} \bibinfo{year}{2021}\natexlab{}.
\newblock \showarticletitle{TRAVART: An Approach for Transforming Variability
  Models}. In \bibinfo{booktitle}{\emph{15th International Working Conference
  on Variability Modelling of Software-Intensive Systems}}
  \emph{(\bibinfo{series}{VaMoS'21})}. \bibinfo{publisher}{ACM},
  \bibinfo{address}{New York, NY, USA}, Article \bibinfo{articleno}{8},
  \bibinfo{numpages}{10}~pages.
\newblock
\showISBNx{9781450388245}


\bibitem[\protect\citeauthoryear{Feichtinger, Sundermann, Th\"{u}m, and
  Rabiser}{Feichtinger et~al\mbox{.}}{2022b}]%
        {feichtingerSPLC2022informationloss}
\bibfield{author}{\bibinfo{person}{Kevin Feichtinger}, \bibinfo{person}{Chico
  Sundermann}, \bibinfo{person}{Thomas Th\"{u}m}, {and} \bibinfo{person}{Rick
  Rabiser}.} \bibinfo{year}{2022}\natexlab{b}.
\newblock \showarticletitle{It's Your Loss: Classifying Information Loss during
  Variability Model Roundtrip Transformations}. In
  \bibinfo{booktitle}{\emph{Proceedings of the 26th ACM International Systems
  and Software Product Line Conference - Volume A}}
  \emph{(\bibinfo{series}{SPLC '22})}. \bibinfo{publisher}{Association for
  Computing Machinery}, \bibinfo{address}{New York, NY, USA},
  \bibinfo{pages}{67–78}.
\newblock
\showISBNx{9781450394437}


\bibitem[\protect\citeauthoryear{Galster, Weyns, Tofan, Michalik, and
  Avgeriou}{Galster et~al\mbox{.}}{2013}]%
        {galster2013variability}
\bibfield{author}{\bibinfo{person}{Matthias Galster}, \bibinfo{person}{Danny
  Weyns}, \bibinfo{person}{Dan Tofan}, \bibinfo{person}{Bartosz Michalik},
  {and} \bibinfo{person}{Paris Avgeriou}.} \bibinfo{year}{2013}\natexlab{}.
\newblock \showarticletitle{Variability in software systems-a systematic
  literature review}.
\newblock \bibinfo{journal}{\emph{IEEE Transactions on Software Engineering}}
  \bibinfo{volume}{40}, \bibinfo{number}{3} (\bibinfo{year}{2013}),
  \bibinfo{pages}{282--306}.
\newblock


\bibitem[\protect\citeauthoryear{Gomaa}{Gomaa}{2005}]%
        {gomaa2005designing}
\bibfield{author}{\bibinfo{person}{Hassan Gomaa}.}
  \bibinfo{year}{2005}\natexlab{}.
\newblock \bibinfo{booktitle}{\emph{{Designing software product lines with
  UML}}}.
\newblock \bibinfo{publisher}{IEEE}.
\newblock


\bibitem[\protect\citeauthoryear{Haugen, W{\k{a}}sowski, and Czarnecki}{Haugen
  et~al\mbox{.}}{2013}]%
        {haugen2013cvl}
\bibfield{author}{\bibinfo{person}{{\O}ystein Haugen}, \bibinfo{person}{Andrzej
  W{\k{a}}sowski}, {and} \bibinfo{person}{Krzysztof Czarnecki}.}
  \bibinfo{year}{2013}\natexlab{}.
\newblock \showarticletitle{{CVL: common variability language}}. In
  \bibinfo{booktitle}{\emph{Proc. of the 17th International Software Product
  Line Conference}}. ACM, \bibinfo{pages}{277--277}.
\newblock


\bibitem[\protect\citeauthoryear{IVML}{IVML}{2015}]%
        {ivmlSpecification2015}
IVML \bibinfo{year}{2015}\natexlab{}.
\newblock \bibinfo{booktitle}{\emph{{Integrated Variability Modeling Language:
  Language Specification}}}.
\newblock \bibinfo{type}{Specification}. \bibinfo{institution}{University of
  Hildesheim}, \bibinfo{address}{Hildesheim, DE}.
\newblock
\urldef\tempurl%
\url{https://projects.sse.uni-hildesheim.de/easy/docs/ivml_spec.pdf}
\showURL{%
\tempurl}


\bibitem[\protect\citeauthoryear{Kang, Cohen, Hess, Novak, and Peterson}{Kang
  et~al\mbox{.}}{1990}]%
        {kang1990feature}
\bibfield{author}{\bibinfo{person}{Kyo~C Kang}, \bibinfo{person}{Sholom~G
  Cohen}, \bibinfo{person}{James~A Hess}, \bibinfo{person}{William~E Novak},
  {and} \bibinfo{person}{A~Spencer Peterson}.} \bibinfo{year}{1990}\natexlab{}.
\newblock \bibinfo{booktitle}{\emph{{Feature-oriented domain analysis (FODA)
  feasibility study}}}.
\newblock \bibinfo{type}{{T}echnical {R}eport}.
  \bibinfo{institution}{Carnegie-Mellon Univ., Pittsburgh, Pa, Software
  Engineering Inst.}
\newblock


\bibitem[\protect\citeauthoryear{Kr\"{u}ger, Nielebock, Krieter, Diedrich,
  Leich, Saake, Zug, and Ortmeier}{Kr\"{u}ger et~al\mbox{.}}{2017}]%
        {beyond_spl}
\bibfield{author}{\bibinfo{person}{Jacob Kr\"{u}ger},
  \bibinfo{person}{Sebastian Nielebock}, \bibinfo{person}{Sebastian Krieter},
  \bibinfo{person}{Christian Diedrich}, \bibinfo{person}{Thomas Leich},
  \bibinfo{person}{Gunter Saake}, \bibinfo{person}{Sebastian Zug}, {and}
  \bibinfo{person}{Frank Ortmeier}.} \bibinfo{year}{2017}\natexlab{}.
\newblock \showarticletitle{Beyond Software Product Lines: Variability Modeling
  in Cyber-Physical Systems}. In \bibinfo{booktitle}{\emph{Proceedings of the
  21st International Systems and Software Product Line Conference - Volume A}}
  \emph{(\bibinfo{series}{SPLC '17})}. \bibinfo{publisher}{Association for
  Computing Machinery}, \bibinfo{address}{New York, NY, USA},
  \bibinfo{pages}{237–241}.
\newblock


\bibitem[\protect\citeauthoryear{Meinicke, Th{\"u}m, Schr{\"o}ter, Benduhn,
  Leich, and Saake}{Meinicke et~al\mbox{.}}{2017}]%
        {FeatureIDE}
\bibfield{author}{\bibinfo{person}{Jens Meinicke}, \bibinfo{person}{Thomas
  Th{\"u}m}, \bibinfo{person}{Reimar Schr{\"o}ter}, \bibinfo{person}{Fabian
  Benduhn}, \bibinfo{person}{Thomas Leich}, {and} \bibinfo{person}{Gunter
  Saake}.} \bibinfo{year}{2017}\natexlab{}.
\newblock \bibinfo{booktitle}{\emph{Mastering Software Variability with
  FeatureIDE}}.
\newblock \bibinfo{publisher}{Springer}.
\newblock


\bibitem[\protect\citeauthoryear{Meixner, Feichtinger, Rabiser, and
  Biffl}{Meixner et~al\mbox{.}}{2022}]%
        {MeixnerVaMos2022EfficientProcessExploration}
\bibfield{author}{\bibinfo{person}{Kristof Meixner}, \bibinfo{person}{Kevin
  Feichtinger}, \bibinfo{person}{Rick Rabiser}, {and} \bibinfo{person}{Stefan
  Biffl}.} \bibinfo{year}{2022}\natexlab{}.
\newblock \showarticletitle{Efficient Production Process Variability
  Exploration}. In \bibinfo{booktitle}{\emph{Proceedings of the 16th
  International Working Conference on Variability Modelling of
  Software-Intensive Systems}} \emph{(\bibinfo{series}{VaMoS '22})}.
  \bibinfo{publisher}{Association for Computing Machinery},
  \bibinfo{address}{New York, NY, USA}, Article \bibinfo{articleno}{14},
  \bibinfo{numpages}{9}~pages.
\newblock
\showISBNx{9781450396042}


\bibitem[\protect\citeauthoryear{Meixner, Rinker, Marcher, Decker, and
  Biffl}{Meixner et~al\mbox{.}}{2021}]%
        {Meixner2021a}
\bibfield{author}{\bibinfo{person}{Kristof Meixner}, \bibinfo{person}{Felix
  Rinker}, \bibinfo{person}{Hannes Marcher}, \bibinfo{person}{Jakob Decker},
  {and} \bibinfo{person}{Stefan Biffl}.} \bibinfo{year}{2021}\natexlab{}.
\newblock \showarticletitle{{A Domain-Specific Language for
  Product-Process-Resource Modeling}}. In \bibinfo{booktitle}{\emph{{IEEE Int.
  Conf. on Emerging Technologies and Factory Automation (ETFA)}}}.
  \bibinfo{publisher}{{IEEE}}.
\newblock


\bibitem[\protect\citeauthoryear{Metzger and Pohl}{Metzger and Pohl}{2014}]%
        {spl_achievement_challenge}
\bibfield{author}{\bibinfo{person}{Andreas Metzger} {and}
  \bibinfo{person}{Klaus Pohl}.} \bibinfo{year}{2014}\natexlab{}.
\newblock \showarticletitle{Software product line engineering and variability
  management: Achievements and challenges}.
\newblock \bibinfo{journal}{\emph{FOSE}}.
\newblock
\urldef\tempurl%
\url{https://doi.org/10.1145/2593882.2593888}
\showDOI{\tempurl}


\bibitem[\protect\citeauthoryear{Pohl, B{\"o}ckle, and van~der Linden}{Pohl
  et~al\mbox{.}}{2005}]%
        {pohl2005software}
\bibfield{author}{\bibinfo{person}{Klaus Pohl}, \bibinfo{person}{G{\"u}nter
  B{\"o}ckle}, {and} \bibinfo{person}{Frank~J van~der Linden}.}
  \bibinfo{year}{2005}\natexlab{}.
\newblock \bibinfo{booktitle}{\emph{{Software Product Line Engineering:
  Foundations, Principles and Techniques}}}.
\newblock \bibinfo{publisher}{Springer Science \& Business Media}.
\newblock


\bibitem[\protect\citeauthoryear{pure-systems GmbH}{pure-systems GmbH}{2023}]%
        {puresystemsUserguide2022}
\bibfield{author}{\bibinfo{person}{pure-systems GmbH}.}
  \bibinfo{year}{2023}\natexlab{}.
\newblock \bibinfo{title}{pure::variants User's Guide}.
\newblock
\newblock
\urldef\tempurl%
\url{https://www.pure-systems.com/fileadmin/downloads/pure-variants/doc/pv-user-manual.pdf}
\showURL{%
\tempurl}
\newblock
\shownote{Version 6.0.1.685, last access 2023-04-06.}


\bibitem[\protect\citeauthoryear{Raatikainen, Tiihonen, and
  M{\"a}nnist{\"o}}{Raatikainen et~al\mbox{.}}{2019}]%
        {Raatikainen2019software}
\bibfield{author}{\bibinfo{person}{Mikko Raatikainen}, \bibinfo{person}{Juha
  Tiihonen}, {and} \bibinfo{person}{Tomi M{\"a}nnist{\"o}}.}
  \bibinfo{year}{2019}\natexlab{}.
\newblock \showarticletitle{Software product lines and variability modeling: A
  tertiary study}.
\newblock \bibinfo{journal}{\emph{Journal of Systems and Software}}
  \bibinfo{volume}{149} (\bibinfo{year}{2019}), \bibinfo{pages}{485--510}.
\newblock


\bibitem[\protect\citeauthoryear{Romano, Feichtinger, Beuche, Ryssel, and
  Rabiser}{Romano et~al\mbox{.}}{2022}]%
        {RomanoMODEVAR2022}
\bibfield{author}{\bibinfo{person}{Dario Romano}, \bibinfo{person}{Kevin
  Feichtinger}, \bibinfo{person}{Danilo Beuche}, \bibinfo{person}{Uwe Ryssel},
  {and} \bibinfo{person}{Rick Rabiser}.} \bibinfo{year}{2022}\natexlab{}.
\newblock \showarticletitle{Bridging the Gap between Academia and Industry:
  Transforming the Universal Variability Language to pure::variants and Back}.
  In \bibinfo{booktitle}{\emph{Proc. of the 5th International Workshop on
  Languages for Modelling Variability (MODEVAR), co-located with SPLC 2022}}.
  \bibinfo{publisher}{ACM}.
\newblock


\bibitem[\protect\citeauthoryear{Schmid, Kr\"{o}her, and El-Sharkawy}{Schmid
  et~al\mbox{.}}{2018}]%
        {ivml_easy_producer}
\bibfield{author}{\bibinfo{person}{Klaus Schmid}, \bibinfo{person}{Christian
  Kr\"{o}her}, {and} \bibinfo{person}{Sascha El-Sharkawy}.}
  \bibinfo{year}{2018}\natexlab{}.
\newblock \showarticletitle{Variability Modeling with the Integrated
  Variability Modeling Language (IVML) and EASy-Producer}. In
  \bibinfo{booktitle}{\emph{Proceedings of the 22nd International Systems and
  Software Product Line Conference - Volume 1}} \emph{(\bibinfo{series}{SPLC
  '18})}. \bibinfo{publisher}{Association for Computing Machinery},
  \bibinfo{address}{New York, NY, USA}, \bibinfo{pages}{306}.
\newblock


\bibitem[\protect\citeauthoryear{Schmid, Rabiser, and Gr{\"u}nbacher}{Schmid
  et~al\mbox{.}}{2011}]%
        {schmid2011comparison}
\bibfield{author}{\bibinfo{person}{Klaus Schmid}, \bibinfo{person}{Rick
  Rabiser}, {and} \bibinfo{person}{Paul Gr{\"u}nbacher}.}
  \bibinfo{year}{2011}\natexlab{}.
\newblock \showarticletitle{A comparison of decision modeling approaches in
  product lines}. In \bibinfo{booktitle}{\emph{Proc. of the 5th International
  Workshop on Variability Modelling of Software-Intensive Systems}}. ACM,
  \bibinfo{pages}{119--126}.
\newblock


\bibitem[\protect\citeauthoryear{Schobbens, Heymans, and Trigaux}{Schobbens
  et~al\mbox{.}}{2006}]%
        {schobbens2006feature}
\bibfield{author}{\bibinfo{person}{Pierre-Yves Schobbens},
  \bibinfo{person}{Patrick Heymans}, {and} \bibinfo{person}{Jean-Christophe
  Trigaux}.} \bibinfo{year}{2006}\natexlab{}.
\newblock \showarticletitle{Feature diagrams: A survey and a formal semantics}.
  In \bibinfo{booktitle}{\emph{Proc. of the 14th IEEE International
  Requirements Engineering Conference}}. IEEE, \bibinfo{pages}{139--148}.
\newblock


\bibitem[\protect\citeauthoryear{She, Lotufo, Berger, W{\k{a}}sowski, and
  Czarnecki}{She et~al\mbox{.}}{2010}]%
        {she2010variability}
\bibfield{author}{\bibinfo{person}{Steven She}, \bibinfo{person}{Rafael
  Lotufo}, \bibinfo{person}{Thorsten Berger}, \bibinfo{person}{Andrzej
  W{\k{a}}sowski}, {and} \bibinfo{person}{Krzysztof Czarnecki}.}
  \bibinfo{year}{2010}\natexlab{}.
\newblock \showarticletitle{{The Variability Model of The Linux Kernel}}. In
  \bibinfo{booktitle}{\emph{Proc. of the 5th International Workshop on
  Variability Modelling of Software-intensive Systems}}. ACM,
  \bibinfo{pages}{45--51}.
\newblock


\bibitem[\protect\citeauthoryear{Sundermann, Feichtinger, Engelhardt, Rabiser,
  and Th{\"u}m}{Sundermann et~al\mbox{.}}{2021}]%
        {SundermannUVLSPLC2021}
\bibfield{author}{\bibinfo{person}{Chico Sundermann}, \bibinfo{person}{Kevin
  Feichtinger}, \bibinfo{person}{Dominik Engelhardt}, \bibinfo{person}{Rick
  Rabiser}, {and} \bibinfo{person}{Thomas Th{\"u}m}.}
  \bibinfo{year}{2021}\natexlab{}.
\newblock \showarticletitle{Yet Another Textual Variability Language? A
  Community Effort Towards a Unified Language}. In
  \bibinfo{booktitle}{\emph{Proc. of the 25th International Systems and
  Software Product Line Conference}}. \bibinfo{publisher}{ACM},
  \bibinfo{address}{Leicester, United Kingdom}.
\newblock


\bibitem[\protect\citeauthoryear{Sundermann, Vill, Th{\"{u}}m, Feichtinger,
  Agarwal, Rabiser, Galindo, and Benavides}{Sundermann et~al\mbox{.}}{2023}]%
        {uvl_levels}
\bibfield{author}{\bibinfo{person}{Chico Sundermann}, \bibinfo{person}{Stefan
  Vill}, \bibinfo{person}{Thomas Th{\"{u}}m}, \bibinfo{person}{Kevin
  Feichtinger}, \bibinfo{person}{Prankur Agarwal}, \bibinfo{person}{Rick
  Rabiser}, \bibinfo{person}{Jos{\'{e}}~A. Galindo}, {and}
  \bibinfo{person}{David Benavides}.} \bibinfo{year}{2023}\natexlab{}.
\newblock \showarticletitle{UVLParser: Extending {UVL} with Language Levels and
  Conversion Strategies}. In \bibinfo{booktitle}{\emph{Proceedings of the 27th
  {ACM} International Systems and Software Product Line Conference - Volume B,
  {SPLC} 2023, Tokyo, Japan}}. \bibinfo{publisher}{{ACM}},
  \bibinfo{pages}{39--42}.
\newblock


\bibitem[\protect\citeauthoryear{Th\"{u}m, Seidl, and Schaefer}{Th\"{u}m
  et~al\mbox{.}}{2019}]%
        {ThuemLanguageLevelsFM2019}
\bibfield{author}{\bibinfo{person}{Thomas Th\"{u}m}, \bibinfo{person}{Christoph
  Seidl}, {and} \bibinfo{person}{Ina Schaefer}.}
  \bibinfo{year}{2019}\natexlab{}.
\newblock \showarticletitle{On Language Levels for Feature Modeling Notations}.
  In \bibinfo{booktitle}{\emph{Proceedings of the 23rd International Systems
  and Software Product Line Conference - Volume B}}
  \emph{(\bibinfo{series}{SPLC '19})}. \bibinfo{publisher}{ACM},
  \bibinfo{address}{New York, NY, USA}, \bibinfo{pages}{158–161}.
\newblock
\showISBNx{9781450366687}


\bibitem[\protect\citeauthoryear{Veer and Dallaway}{Veer and Dallaway}{2011}]%
        {veer2011ecos}
\bibfield{author}{\bibinfo{person}{Bart Veer} {and} \bibinfo{person}{John
  Dallaway}.} \bibinfo{year}{2011}\natexlab{}.
\newblock \bibinfo{title}{The eCos Component Writer’s Guide}.
\newblock \bibinfo{howpublished}{Manual, available online at
  http://www.gaisler.com/doc/ecos-2.0-cdl-guide-a4.pdf}.
\newblock


\end{thebibliography}

\end{document}